\documentclass{iaus}
\usepackage{graphicx}
\usepackage{sidecap}

\newcommand{\EQ}{\begin{equation}}
\newcommand{\EN}{\end{equation}}
\newcommand{\EQA}{\begin{eqnarray}}
\newcommand{\ENA}{\end{eqnarray}}

\newcommand{\EEq}[1]{Equation~(\ref{#1})}
\newcommand{\Eq}[1]{Eqs.~(\ref{#1})}

\newcommand{\Sec}[1]{Section~\ref{#1}}

\newcommand{\Fig}[1]{Fig.~\ref{#1}}

\newcommand{\bra}[1]{\langle #1\rangle}

{}
{}
{}

{}
{}
{}
{}
{}
{}
{}
{}
{}
{}
{}
{}
{}
{}
{}
{}
{}
{}

{}
{}
{}

{}

{}
{}

\newcommand{\UU}{\mbox{\boldmath $U$} {}}

\newcommand{\BB}{\mbox{\boldmath $B$} {}}

\newcommand{\JJ}{\mbox{\boldmath $J$} {}}

\newcommand{\AAA}{\mbox{\boldmath $A$} {}}

\newcommand{\ff}{\mbox{\boldmath $f$} {}}

\newcommand{\FF}{\mbox{\boldmath $F$} {}}

\newcommand{\nab}{\mbox{\boldmath $\nabla$} {}}

\newcommand{\erf}{{\rm erf}}

\newcommand{\DD}{{\rm D} {}}

\newcommand{\const}{{\rm const}  {}}

\def\Rm{\mbox{\rm Re}_M}

\def\cs{c_{\rm s}}

\def\kf{k_{\rm f}}

\def\half{{\textstyle{1\over2}}}

\newcommand{\yapj}[3]{ #1, {ApJ,} {#2}, #3}

\newcommand{\yapjl}[3]{ #1, {ApJ,} {#2}, #3}

\newcommand{\yana}[3]{ #1, {A\&A,} {#2}, #3}

\newcommand{\ybook}[3]{ #1, {#2} (#3)}

\title[Dynamo generated plasmoid ejections] 
{Dynamo generated field emergence through recurrent plasmoid ejections}

\author[J{\"o}rn Warnecke \& Axel Brandenburg]   
{J{\"o}rn Warnecke$^{1,2}$
 \and Axel Brandenburg$^{1,2}$}

\affiliation{$^1$Nordita, AlbaNova University Center, \\Roslagstullsbacken 23,
SE-10691 Stockholm, Sweden \\email: {\tt joern@nordita.org} \\[\affilskip]
$^2$Department of Astronomy, AlbaNova University Center, \\Stockholm University, 
SE 10691 Stockholm, Sweden}

\pubyear{2011}
\volume{273}  
\pagerange{119--126}
\jname{Physics of Sun and Star Spots}
\editors{D.P. Choudhary \& K.G. Strassmeier, eds.}
\begin{document}

\maketitle

\begin{abstract}
Magnetic buoyancy is believed to drive the transport of magnetic flux tubes 
from the 
convection zone to the surface of the Sun. The magnetic fields form twisted 
loop-like structures in the solar atmosphere.
In this paper we use helical forcing to produce a large-scale
dynamo-generated magnetic field, which rises even without magnetic
buoyancy.
A two layer system is used as computational domain where the
upper part represents the solar atmosphere. Here, the evolution 
of the magnetic field is solved with the stress--and--relax method.
Below this region a magnetic field is produced by a helical forcing function 
in the momentum equation, which leads to dynamo action.
We find twisted magnetic fields emerging frequently to the outer layer, forming
arch-like structures. In addition, recurrent plasmoid ejections can be found by 
looking at space--time diagrams of the magnetic field.
Recent simulations in spherical coordinates show similar results.

\end{abstract}

\firstsection 
\section{Introduction}
 
The solar magnetic field is broadly believed to be in the form of concentrated
flux ropes 
in the bulk of the convection zone. At the solar surface they emerge to 
form bipolar regions and sunspots in the photosphere that appear as twisted loop-like 
structures in the higher atmosphere.
However, there is no clear 
evidence that magnetic fields are generated in flux tubes that emerge from the 
tachocline all the way to the surface of the Sun. Numerical simulations have 
successfully shown that magnetic buoyancy, which has been thought to be the main 
driver of flux tube emergence, can be efficiently suppressed by downward 
pumping due to the stratification with concentrated downdraft in the solar 
convection zone (Nordlund et al.\ 1992; Tobias et al.\ 1998). Large-scale 
dynamo simulations suggest that flux tubes 
are primarily a feature of the kinematic dynamo regime, but tend to be 
less pronounced in the nonlinear stage (K\"apyl\"a et al.\ 2008). An 
alternative mechanism might simply be the relaxation of strongly twisted 
magnetic fields reaching the surface of the Sun. Twisted magnetic fields are 
produced by a large-scale dynamo mechanism which is generally believed to be the
source of solar magnetic activity (Parker 1979). In order to study the emergence of 
helical magnetic fields from a dynamo, we consider a model that combines a 
direct simulation of a turbulent large-scale dynamo with a simple treatment for
the evolution of nearly force-free magnetic fields above the surface of the 
dynamo. In the context of force-free magnetic field extrapolations this method 
is also known as the stress--and--relax method (Valori et al.\ 2005).
Including a nearly force-free field in the upper part of the domain has the
additional benefit of allowing a more realistic modeling of the dynamo itself.
This is important, because it is known that the properties of the
large-scale magnetic field depend strongly on the boundary conditions. 
In the upper atmosphere, direct numerical simulation of the solar corona 
show force-free magnetic fields (Gudiksen \& Nordlund 2005).

Above the solar surface, we expect helical magnetic fields to drive flares 
and coronal mass ejections through the Lorentz force.
In the present paper we highlight some of the main results of our earlier
work (Warnecke \& Brandenburg 2010) and present recent applications and 
results using spherical coordinates

\section{The Model}
\label{model}
A two layer system is used, where the upper part is modelled 
as a nearly force-free magnetic field 
by using the stress--and--relax method (Valori et al.\ 2005)
and in the lower part a dynamo field is generated through helically
forced turbulence.
We combine these two layers by simply turning off 
terms that should not be included in the upper part of the domain.
We do this with error function profiles of the form
\begin{equation}
\theta_w(z)=\half\left(1-\erf{z\over w}\right),
\end{equation}
where $w$ is the width of the transition.

\subsection{Stress--and--relax method}

The equation for the velocity in the stress--and--relax method is
similar to the usual momentum equation,
except that there is neither pressure, nor gravity, nor other driving forces on
the right-hand side, so we just have
\begin{equation}
{\DD\UU\over\DD t}=\JJ\times\BB/\rho+\FF_{\rm visc},
\label{DUDtext}
\end{equation}
where $\JJ\times\BB$ is the Lorentz force,
$\JJ=\nab\times\BB/\mu_0$ is the current density,
$\mu_0$ is the vacuum permeability,
$\FF_{\rm visc}$ is the viscous force,
and $\rho$ is here treated as a constant
that determines the strength of the velocity correction.
\EEq{DUDtext} is solved together with the uncurled induction equation,
\begin{equation}
{\partial\AAA\over\partial t}=\UU\times\BB+\eta\nabla^2\AAA,
\label{DDBB}
\end{equation}
with $\eta$ being the magnetic diffusivity.

\subsection{Forced dynamo region}

In the lower part the velocity is driven by a forcing function and the density
is evolved using the continuity equation,
\begin{equation}
{\DD\UU\over\DD t}=-\nab h+\ff +\JJ\times\BB/\rho+\FF_{\rm visc},\quad\quad
{\DD h\over\DD t}=-c_s^2\nab\cdot\UU,
\label{DDUU}
\end{equation}
where $\FF_{\rm visc}$ is the viscous force,
$h=\cs^2\ln\rho$ is the specific pseudo-enthalpy,
$\cs=\const$ is the isothermal sound speed,
and $\ff$ is a forcing function that drives turbulence in the interior
and consists of random 
plane helical transversal waves with an average forcing wavenumber $\kf$.
The pseudo-enthalpy $h$ is given by $\rho^{-1}\nab p=\cs^2\nab\ln\rho=\nab h$.
Equations~(\ref{DDUU}) are solved together with the induction \Eq{DDBB}.

The simulation box is horizontally periodic.
For the magnetic field
we adopt vertical-field and perfect-conductor conditions
at the top and bottom boundaries, respectively.
For the velocity we employ stress-free conditions at both boundaries.
In this paper we present direct numerical simulations using the 
{\sc Pencil Code}\footnote{\texttt{http://pencil-code.googlecode.com}},
a modular high-order code (sixth order in space and third-order
in time) for solving a large range of partial differential
equations
\section{Results}
\begin{figure}[t!]\begin{center}
\includegraphics[width=7cm]{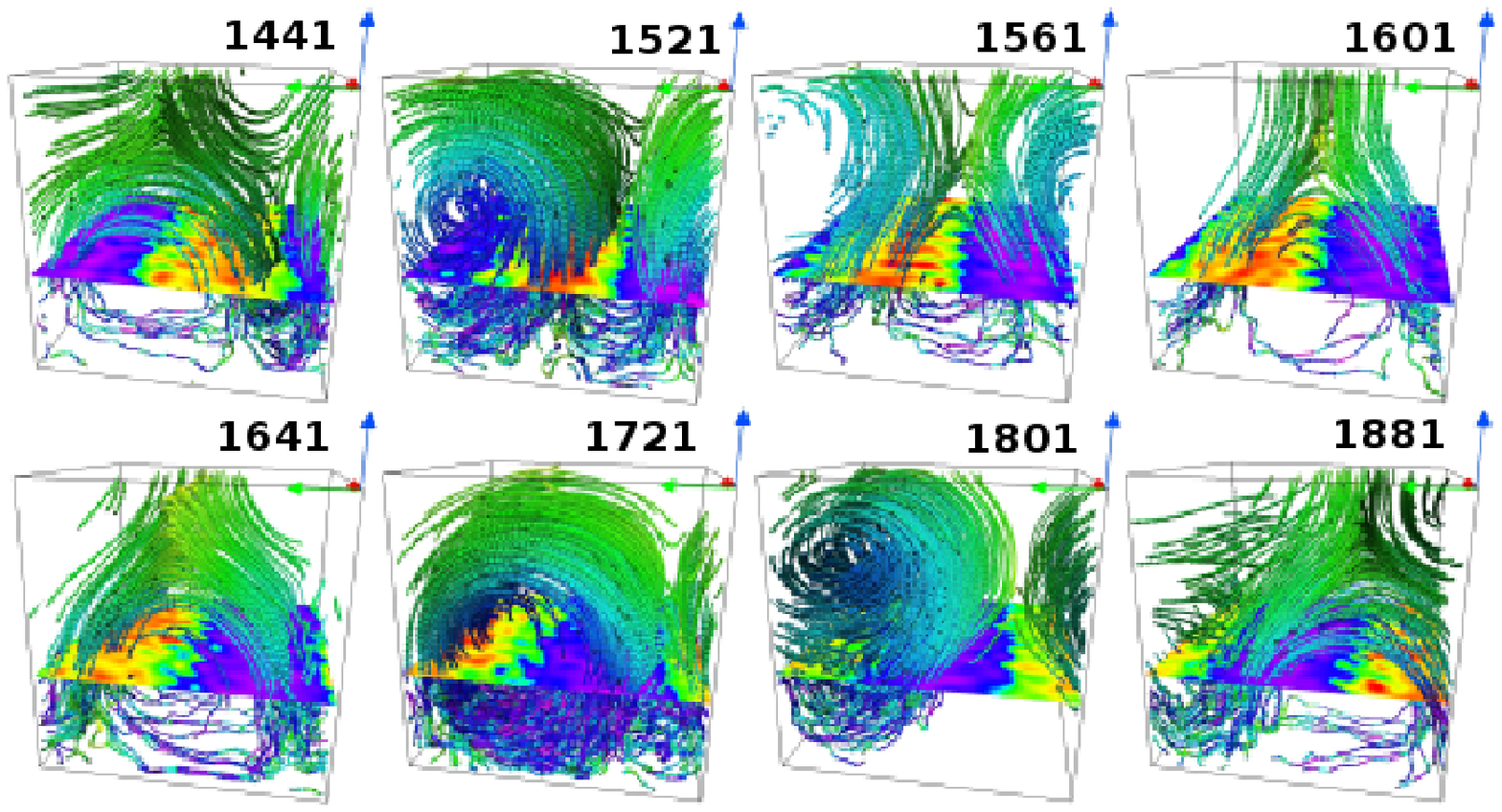}
\includegraphics[width=6.3cm]{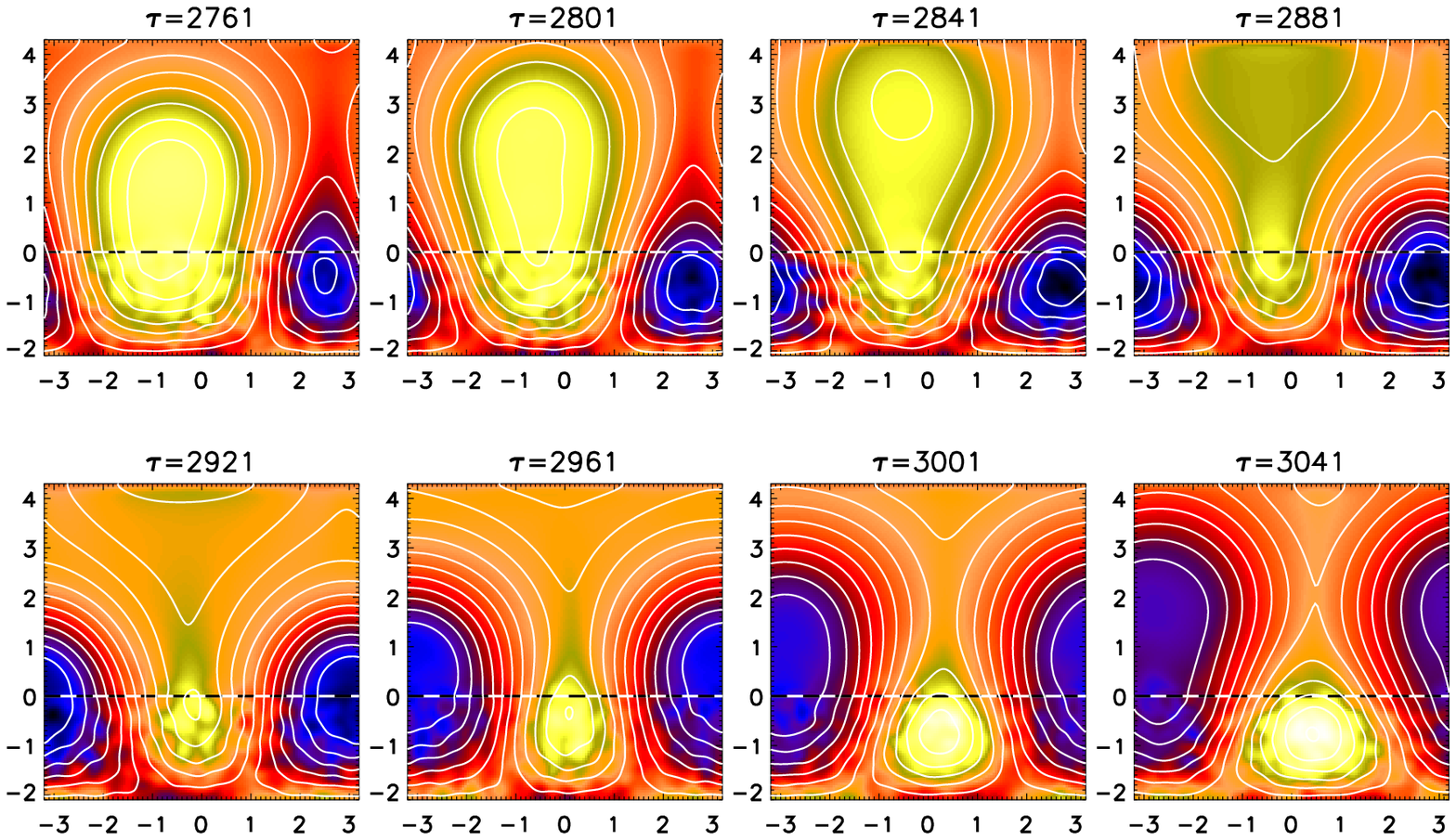}
\end{center}\caption[]{
{\it Left panel}: Time series of arcade formation and decay. Field lines are colored by 
their local field strength which increases from purple to green.
The inclined plane in the box shows $B_z$ increasing from red (positive) to purple (negative).
The normalized time $\tau$ is giving in each panel.
{\it Right panel}: Time series of the formation of a plasmoid ejection. Contours of 
$\bra{A_x}_{x}$ are shown together with a color-scale representation of $\bra{B_x}_x$;
dark blue stands for negative and red for positive values.
The contours of $\bra{A_x}_{x}$ correspond to field lines of $\bra{\BB}_x$
in the $yz$ plane. The dotted horizontal lines show the location of the surface at $z=0$.
Adapted from Warnecke \& Brandenburg (2010).
}
\label{AI}
\end{figure}

\begin{figure}[t!]
\begin{center}
\includegraphics[width=6.7cm]{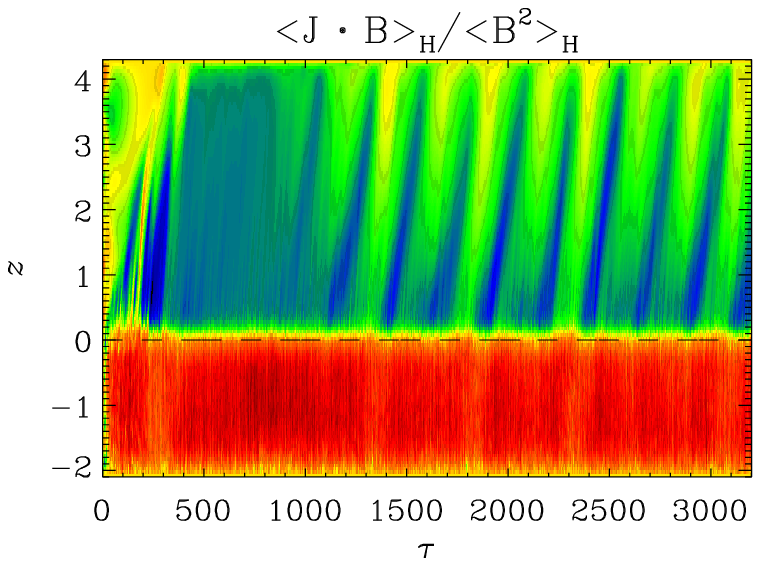}
\includegraphics[width=6.7cm]{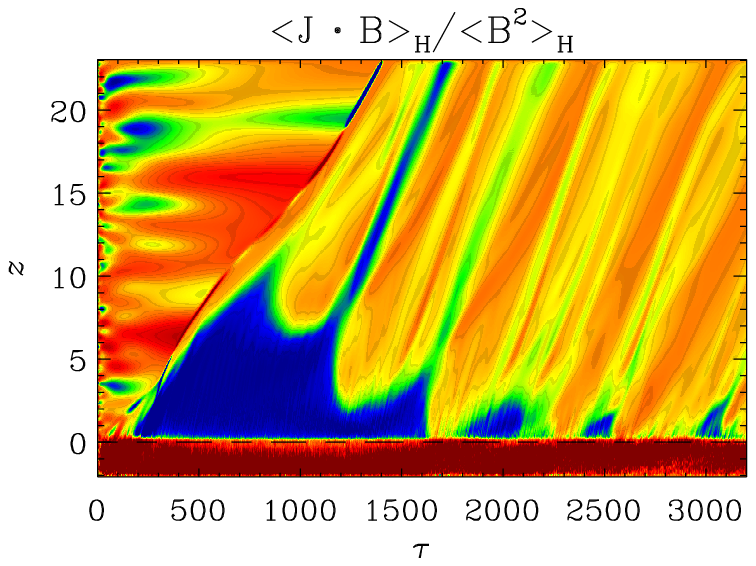}
\end{center}\caption[]{
{\it Left panel}: Dependence of $\bra{\JJ\cdot\BB}_{\rm H}/\bra{\BB^2}_{\rm H}$
versus time $\tau$ and height $z$ for $L_z=6.4$ with $\Rm=3.4$
{\it Right panel}: Similar to the left panel, but for $L_z=8\pi$ and $\Rm=6.7$
.
Adapted from Warnecke \& Brandenburg (2010).
}
\label{jb}
\end{figure}

After a period of exponential growth, the magnetic field saturates at $78\%$
of the equipartition field strength, $B_{\rm eq}$ in the turbulent
layer. This behavior is typical for forced dynamo action. The
structure of the magnetic field is a large-scale field in the
turbulent zone.
It always shows a systematic variation in one of the two horizontal
directions.
It is a matter of chance whether this variation is in the $x$ or in the $y$
direction.
After the saturation phase the magnetic field extends well into the upper layer
where it tends to produce an arcade-like structure, as seen in the left panel
of \Fig{AI}. The arcade opens up in the middle above the line where the vertical field 
component vanishes at the surface. This leads to the formation of anti-aligned field 
lines with a current sheet in the middle.
The dynamical evolution is clearly seem in a sequence of field line images
in the left hand panel 
of \Fig{AI}, where anti-aligned vertical field lines reconnect above the neutral line 
and form a closed arch with plasmoid ejection.
This arch then changes its 
connectivity at the foot points in one of the two horizontal directions (here the $y$ direction), 
making the field lines bulge upward to produce a new reconnection site with anti-aligned 
field lines some distance above the surface.
Field line reconnection is best seen for two-dimensional magnetic fields, because it is 
then possible to compute a flux function whose contours correspond to field lines in the 
corresponding plane.
In the present case the large-scale component of the
magnetic field varies little in the
$x$ direction, so it makes sense to visualize the field averaged over $x$.
In order to demonstrate that plasmoid ejection is a recurrent phenomenon,
it is convenient to look at the evolution of the ratio
$\bra{\JJ\cdot\BB}_{\rm H}/\bra{\BB^2}_{\rm H}$ versus $t$ and $z$.
This is done in right panel of \Fig{jb} for $L_z=6.4$ and $\Rm=3.4$
and in the left panel of \Fig{jb} for $L_z=8\pi$ and $\Rm=6.7$.
It turns out that in both cases the typical speed of plasmoid
ejecta is about half the rms velocity of the turbulence in the
interior region.

\begin{figure}[t!]\begin{center}
\includegraphics[width=3.3cm]{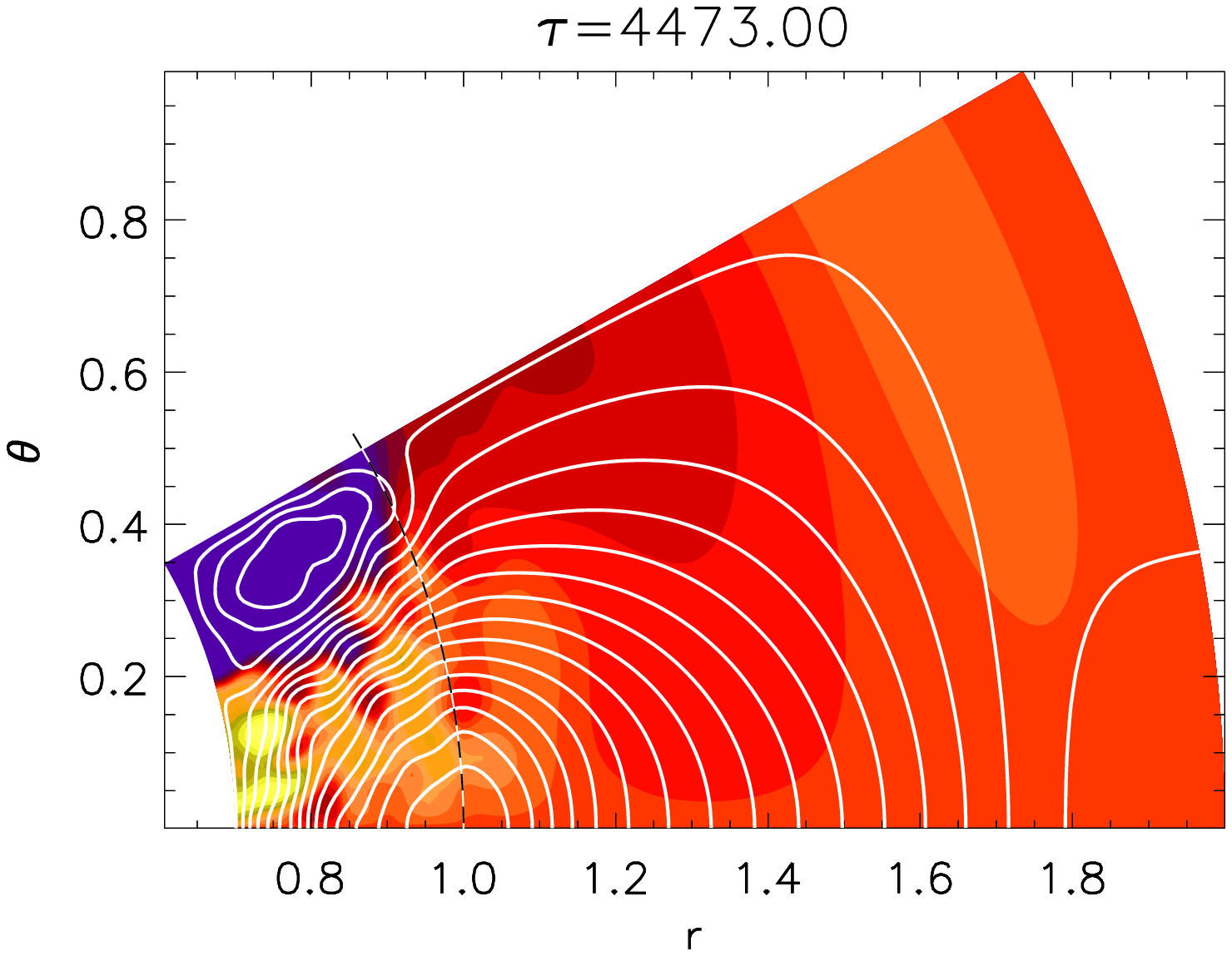}
\includegraphics[width=3.3cm]{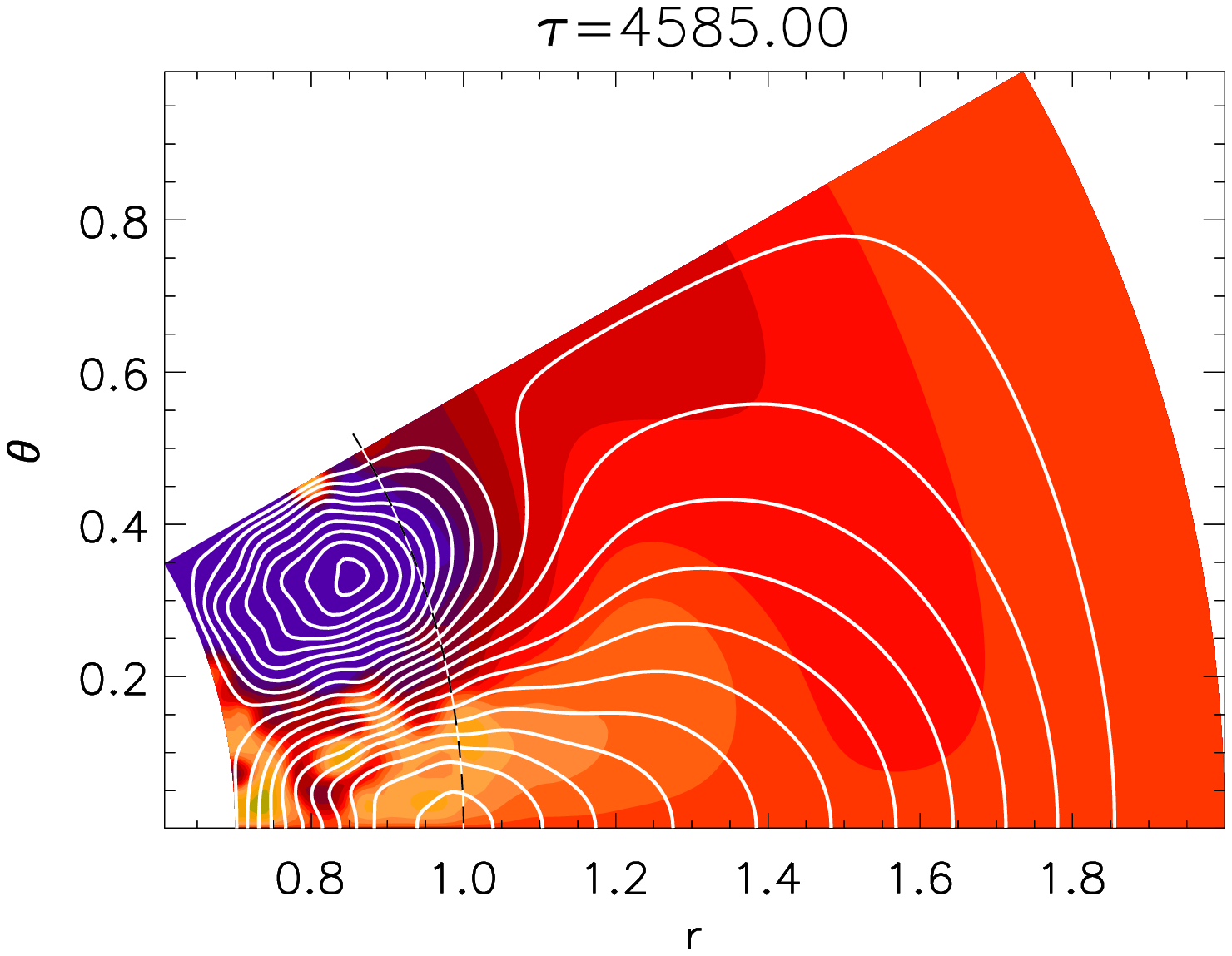}
\includegraphics[width=3.3cm]{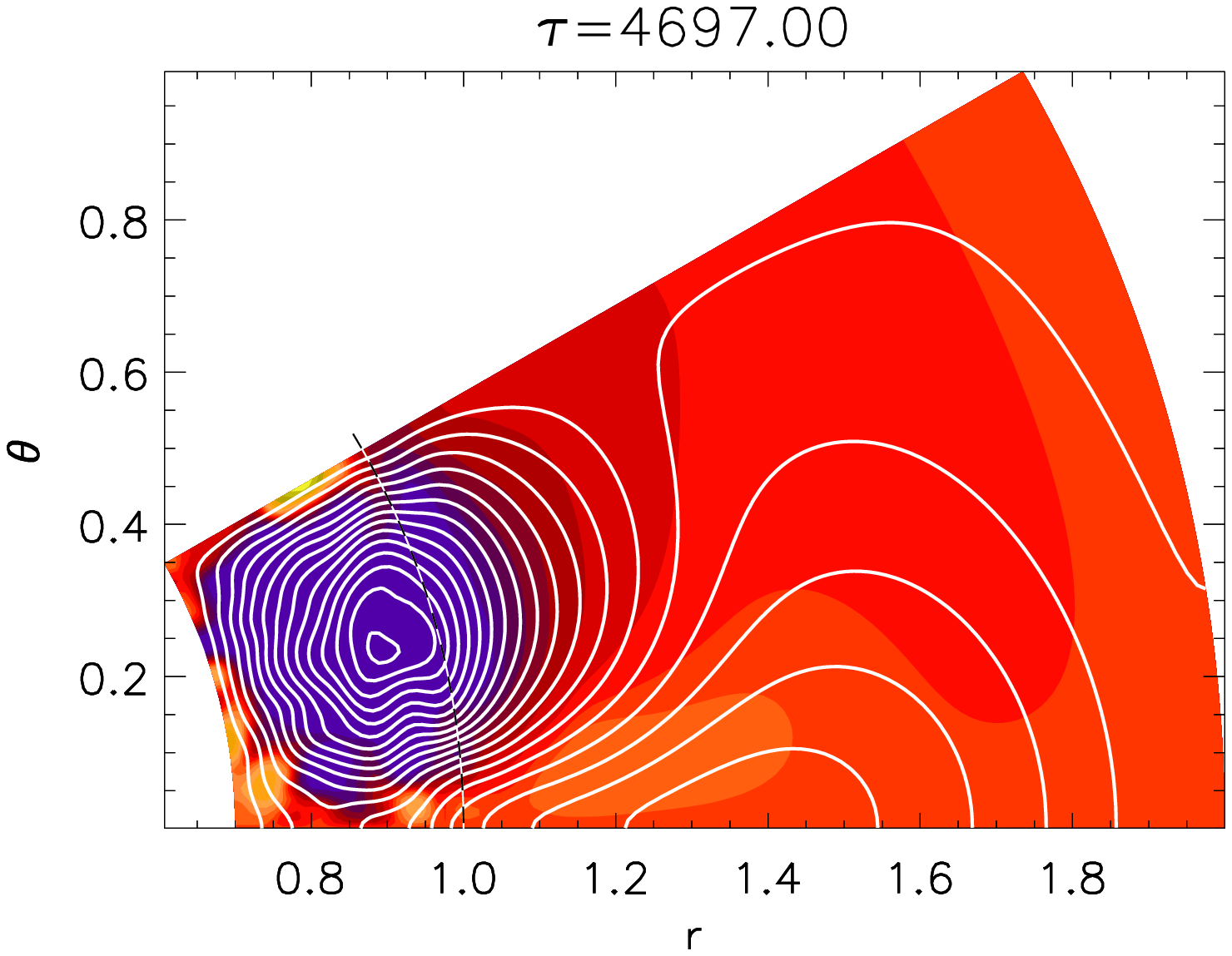}
\includegraphics[width=3.3cm]{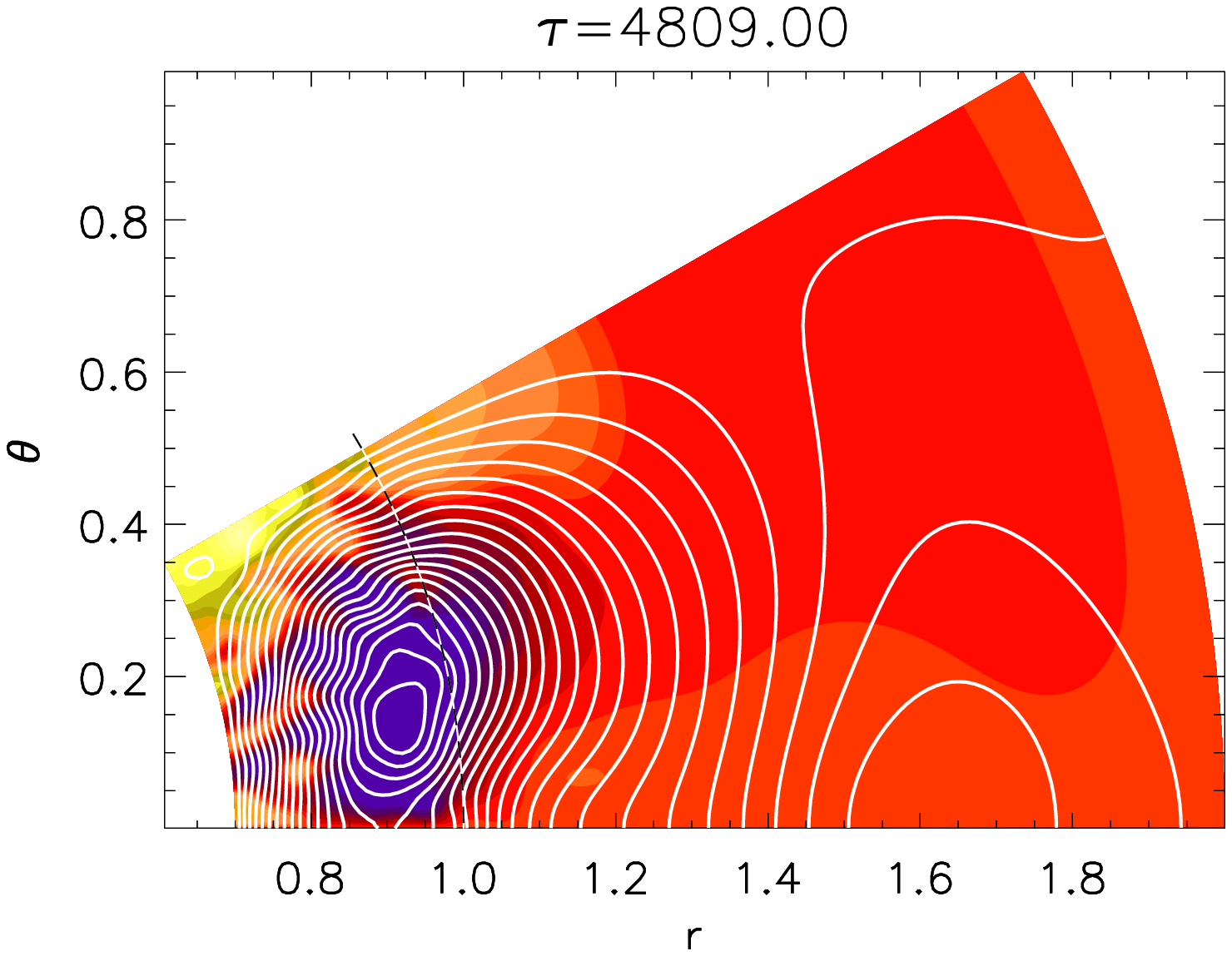}
\end{center}\caption[]{Time series of flux emergence in spherical coordinates. Contours of 
$r\sin{\theta}\bra{A_{\phi}}_{\phi}$ are shown together with a color-scale representation of $\bra{B_{\phi}}_{\phi}$;
dark blue stands for negative and red for positive values.
The contours of $r\sin{\theta}\bra{A_{\phi}}_{{\phi}}$ correspond to field lines of $\bra{\BB}_{\phi}$
in the $r\theta$ plane. The dotted horizontal lines show the location of the surface at $r=1$
solar radii.}
\label{A_sph}
\end{figure}

As an example of further work in this direction, we also present in this paper
magnetic flux emergence in spherical coordinates.
The dynamical evolution can be seen in \Fig{A_sph}, where
a modulated slice covers the convection zone from $0.7$
solar radii through the upper atmosphere to two solar radii.
This meridional slice of a sphere consists, like
the simulation box above, of two layers which contain the same
physical properties.
We solve the same equations as described in \Sec{model}.
Again, there is flux emerge through the surface
above the turbulence zone. This can be seen as a recurrent
event.
Unfortunately, reconnection, current sheets and plasmoid ejections
have not been seen in the present setup, although there are indications
that they do occur in even more recent spherical models where gravity and
density stratification are included.

Our first results are promising in that the dynamics of the magnetic
field in the exterior is indeed found to mimic open boundary conditions
at the interface between the turbulence zone and the exterior at $z=0$.
In particular, it turns out that a twisted magnetic field generated
by the helical dynamo beneath the surface is able to produce flux emergence
in ways that are reminiscent of that found in the Sun. The first
results in spherical coordinates show recurrent flux emergence,
but plasmoid ejections in a curved environment may only be possible
if gravity and density stratification are included.

\end{document}